# Development of a Real-Time Software-Defined Radio GPS Receiver Exploiting a LabVIEW-based Instrumentation Environment

Erick Schmidt, *Student Member, IEEE,* David Akopian, *Senior Member, IEEE,* Daniel J. Pack*, Senior Member, IEEE*

*Abstract*—The ubiquitousness of location based services (LBS) has proven effective for many applications such as commercial, military, and emergency responders. Software-defined radio (SDR) has emerged as an adequate framework for development and testing of global navigational satellite systems (GNSS) such as the Global Position System (GPS). SDR receivers are constantly developing in terms of acceleration factors and accurate algorithms for precise user navigation. However, many SDR options for GPS receivers currently lack real-time operation or could be costly. This paper presents a LabVIEW (LV) and C/C++ based GPS L1 receiver platform with real-time capabilities. The system relies on LV acceleration factors as well as other C/C++ techniques such as dynamic link library (DLL) integration into LV and parallelizable loop structures, and single input multiple data (SIMD) methods which leverage host PC multi-purpose processors. A hardware testbed is presented for compactness and mobility, as well as software functionality and data flow handling inherent in LV environment. Benchmarks and other real-time results are presented as well as compared against other state-of-the-art open-source GPS receivers.

*Index Terms*---global navigation satellite systems (GNSS), global positioning systems (GPS), software defined radio (SDR), real-time receiver, acceleration factors.

## I. Introduction

THE success of US Global Positioning System (GPS) in enabling various location-based services triggered extensive studies in related positioning methods, baseband technologies, mitigation of errors and interference. Other similar systems referred to as Global Navigation Satellite Systems (GNSS) have been also deployed using similar signaling methods and infrastructure solutions [1], [2]. Availability of accurate source of user position, velocity, and time (PVT) significantly impacted other technologies such as wireless communication, military equipment, transportation, etc.

Conventional GPS receivers operate in open-sky environments, and are challenged by signal blockages inside buildings, urban canyons, and underground. Also signal interferences and spoofing may deny GPS availability and disrupt the operation of other systems which rely on GPS data. Extensive engineering effort is directed in addressing these challenges to achieve more robust operation of the receivers and expand their coverage to as many denied areas as possible.

Serious advances are achieved in increasing sensitivities of the receivers which have access to terrestrial signaling channels. Using communication links, one can retrieve orbital data of satellites, and receiver coarse position and time estimates from wireless networks which significantly helps in enhancing the operation of GPS receivers. This approach is called Assisted GPS (A-GPS) [1]-[3] and is estimated to improve start-up sensitivity by as much as 20 dB when used in combination with advanced parallel correlators. The A-GPS is standardized for telecommunication networks in terms of defining logistics of communicating various assistance data, and computing-delivering PVT information. It is also recommended by FCC E911 mandate as a solution that will assist emergency services. Many advances are also made for better immunity of GPS receivers against interferences and spoofing [4]-[7].

Despite this progress, GPS operation is still denied in many indoor and other weak signal environments. Also, despite many reported spoofing mitigation methods, these interference techniques also evolve and bring new challenges for GPS equipment. The researchers thrive to improve the performance of the receivers, address continuously evolving spoofing threats, and explore new transformative concepts, and they need proper instrumentation and software to support their efforts. As such, software-defined radio (SDR) solutions become popular because of providing full control of receiver operations, so the researchers can integrate and test their methods without redesigning all receiver chains.

SDR solutions provide extended capabilities for tightly coupled integrations. In this context, SDR integrations provide research capabilities for many purposes such as indoor and vehicular navigation in GPS-denied areas. Proposed integrations include GPS and magnetic positioning systems (MPS) [8] tight coupling via an SDR system. Other SDR solutions provide access to tracking correlators for multipath studies in urban canyons on vehicular applications [9]. Ranging GPS-like radios have been explored as well for indoor positioning applications involving time-of-arrival (TOA) measurements combined with receiver signal strength (RSS)

WLAN-type measurements [10]. In this context, SDR solutions perfectly fit and can be used for such systems. Inertial navigation system (INS) integrations with ranging sensors and GPS-like radios become attractive for SDR fusion applications as well [11]-[14].

SDRs receive sampled data from peripheral RF front-ends and apply signal processing using general purpose computing resources such as computer processors and general-purpose accelerators such as DSPs and FPGAs. The GPS SDRs are currently available in various formats. Proprietary commercial solutions are common and some of them provide application programming interface (API) access to selected modules for 3rd party revisions of these modules [15]. They are typically implemented as C/C++ solutions to provide high-quality and robust performance. The drawback of these solutions is in constraining access to limited functionalities of the receiver. Standalone open-source or open-reported C/C++ solutions are also common [16] and provide full receiver control. The operation quality is not typically guaranteed, the user interfaces are basic, and front-end compatibility is limited. Another interesting SDR category exploits rich library support of dedicated frameworks such as GNU radio [17]. This concept provides essential modular support of integrating available SDR components for fast prototyping, supports many front-ends including popular USRP front-ends [18]. Its excellent development environment but requires somewhat longer environment learning period compared to other concepts. MATLAB/Simulink-based solutions reduce development cycle and are convenient for research studies, but they are typically not real-time [2] or limited to basic receiver grades. Recently basic-grade academic GPS SDR solutions exploiting hybrid development environments, such as C/C++ libraries integrated into LabVIEW (LV) are reported in [19]-[21]. The SDR computing platforms and accelerators such as FPGAs and DSPs combined with multi-purpose processors have also gained attention [20]-[23]. Other solutions exploiting graphical processing unit (GPU) on host PCs for massive parallel processing operations were reported [24] in addition to C/C++ features reported in [9].

This paper presents a new generation of LV-based GPS receivers that achieves high performance operation exploiting multiple strengths of LV environment. It describes and characterizes the impact of using various inherent LV mechanisms, such as multithreading, parallelization, and dedicated loop-structures. It exploits C/C++ optimization techniques for single instruction multiple data (SIMD) capable processors in software correlators for real-time operation of GNSS tracking loops, among other acceleration factors. It also provides comparative analysis of LV-based receivers with *GNU-Radio* and representative open-source C/C++ solutions such as *GNSS-SDR* [25]. This paper also discusses performance comparison metrics to assess general capabilities and robustness of the receiver. It is demonstrated that LV-based features provide competitive real-time solutions for fast prototyping of receiver algorithms. The described GPS SDR platform is a competitive solution capable of using advanced user interface and visualization LV libraries in real-time operation. The proposed receiver also exploits modularity on SDRs by splitting GPS functionality into three main components: acquisition, tracking, and navigation. This paper will explain the proposed GPS L1 SDR receiver testbed and its implementation in detail in the following sections.

The paper is organized as follows. Section II will describe conventional GPS functionality modules. Section III will explain the overall system architecture such as hardware and software components that were used and chosen based on compactness and mobility. Section IV will focus on acceleration factors as well as SDR configurability options for real-time receiver operation. Section V will discuss receiver performance metrics with respect to acquisition, tracking, and navigation modules, as well as comparisons with open-source solutions such as *fast-gps* [16] and *GNSS-SDR* [25]. Section VI will present conclusion remarks.

## II. CONVENTIONAL GPS BASEBAND MODULES

The main communication system focus on this paper is on the conventional L1 civilian GPS signal: a direct sequence spread-spectrum (DSSS) signal consisting of a main binary phase shift keying (BPSK) navigation payload signal operating at 50 Hz, spread by a faster rate BPSK pseudorandom code signal (PRN) at 1.023 Mchips/sec. The spreading sequence is called the coarse acquisition (C/A) code. Finally, the signal is mixed to a sinusoidal carrier at 1.57542 GHz corresponding to the L1 band, forming a band-pass signal transmitted by GPS satellites orbiting the Earth, for civilian positioning.

To successfully demodulate GPS L1 signals and receive navigation data, the receiver synchronizes to the incoming signal by correlating to incoming PRN code and wiping off the L1 carrier. To do this, the receiver generates local replicas of the code (time) and carrier (frequency), and mixes them with the received signal. Therefore, synchronization is achieved in both frequency and time domains, by estimating the Doppler modulation called carrier-phase, and by aligning the signal and PRN replica to estimate their relative shift, called code-phase, respectively. The synchronization is typically performed in two phases: acquisition (coarse) and tracking (fine). After fine synchronization, navigation data are simply obtained after sinusoidal carrier and PRN code wipe-off. Also, to estimate user position, synchronization should be very precise to compute distance travelled from incoming signal.

*1) Conventional and advanced acquisition baseband module*

The first stage of baseband signal processing is acquisition (coarse synchronization) of GPS satellites. Conventional receivers achieve acquisition by searching over a two-dimensional time-frequency discrete zone for each satellite. The receiver replicates candidate PRN code and residual carrier modulation pairs, attempting to match those of the received signal. Several signal replica candidates are locally generated and correlated with incoming signal to find a match and thus to identify input parameters. The correlation operation assumes an element-wise multiplication of the received samples with the samples of each replica. Then resulting products are added over coherent integration lengths, by exploiting periodicity of GPS PRN codes. Typically, a threshold is applied to the correlation and integration result, to determine if a signal acquisition has been reached. If a certain PRN is acquired, then this coarse synchronization is terminated, and the receiver starts the tracking (fine synchronization) stage for that satellite; if not, the search continues and moves to the next code-phase/frequency option for each satellite PRN.

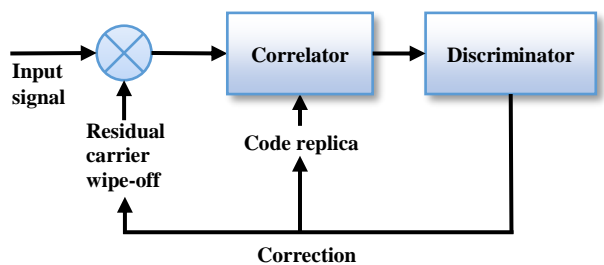

Fig. 1. A simplified tracking feedback loop for a single channel.

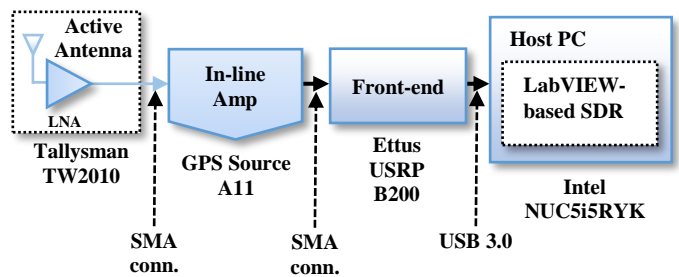

Fig. 2. Hardware architecture for proposed SDR.

Conventional GPS baseband algorithms such as coarse acquisition could be computationally costly for real-time operation due to its nature of typically three-dimensional search space: code-phase, Doppler frequency, and satellite number. At the same time several optimizations can be achieved by means of fast-Fourier transform (FFT) implementations and other joint algorithm computations [16]. An extended parallel code-phase (PCS) search algorithm which leverages its modular structure for concurrent joint search-space in code-phase and frequency is used in the proposed SDR [26]. The algorithm implements a massive correlator by concurrently combining code-phase and Doppler frequency search bins while sharing computations even for different satellites as the received signal forward FFT output is reused for all iterations. The algorithm is effectively implemented in C/C++ language and improves massively on speed when compared to conventional FFT-based PCS methods [20]. Comparative results showing algorithmic acceleration factors will be discussed in later sections.

*2) Tracking loops*

Once a channel has been acquired, a fine synchronization to keep track of the candidate channel is desired. Fig. 1 shows a common tracking feedback loop used for GPS signals. Similar to acquisition, but now in a finer search grid, tracking loops use closed loops to continuously follow the PRN code and carrier parameters of current channel. To determine code and carrier changes of incoming signal, conventional feedback loops such as delay lock loop (DLL) for code-phase estimation, phase lock loop (PLL) and frequency lock loop (FLL) for Doppler modulations, are implemented [2]. Once these loops obtain correction measurements, the discriminator processes correlator outputs to provide measurable quantities which are used as feedback for next iteration, therefore achieving a continuous lock to the incoming signal (see Fig. 7).

*3) Navigation module*

Well-known navigation algorithms discussed in literature [2], [16], are implemented in the receiver such as least-squares (LS) methods and position averaging.

### III. SYSTEM ARCHITECTURE

The development and testbed platform is implemented at the Software Communications and Navigation Systems (SCNS) Laboratory at the University of Texas at San Antonio (UTSA). This paper provides detailed description of a real-time LV-based SDR receiver with various applied concepts to enable high-performance computing. The testbed also includes a GPS simulator which acts as a generator/transmitter of GPS signals.

Current implementation is using National Instruments (NI) GNSS simulation toolkit [27] that can simulate different satellites, signals strengths, Doppler effects, user movement, among other features.

The SDR is a novel LV-based GPS receiver which implements GPS L1 baseband processing functionalities into C/C++ software components that have been compiled as dynamic link libraries (DLLs). It exploits several fast algorithms, LV-based acceleration factors, C/C++ parallel algorithms such as SIMD matching on multi-purpose host CPUs, and external C/C++-based application program interface (API) optimized libraries, among other features. This section will describe the overall development hardware and software testbed components used in this SDR system and their specifications as well as the overall receiver architecture and LV functionalities.

*A. Hardware components*

The hardware used in the system was chosen with an effort to achieve portability and mobility for the SDR receiver. Fig. 2 shows exemplified hardware components and their connectivity. The hardware of choice is an Ettus USRP B200 [18] front-end along with an Intel NUC 5i5RYK [28] serving as the host PC where the baseband software resides. Table I shows specifications for the host PC. The host PC is a small form-factor device. The processor is a low-power multi-purpose CPU (it consumes 15 W). An active antenna, Talisman TW2010 [29], which is power-driven by an in-line amplifier, model A11 from GPS Source [30], are the RF components connected to the front-end. The front-end is fed via an USB 3.0 for power as well as interface with the host PC (A USB Y-cable is used to pull extra power). The front-end RF coverage is broad and can process various GNSS signals including GPS L1 band (1575.42 MHz) that is used for the performance evaluation of the platform. The internal analog-to-digital (ADC) sampling rate from the front-end can achieve speeds up to 56 Msps. Since the GPS SDR system is tested for 5 MHz sampling rate in I-Q interleaved format, the total throughput through the USB cable can reach up to 10 MB/s for INT8 data format, well suitable for USB 3.0 capabilities and with enough resolution for GPS samples.

TABLE I
INTEL NUC HOST PC DEVICE SPECIFICATIONS

| Host PC | Intel NUC5i5RYK |
|---|---|
| CPU | Intel Core i5-5250U @ 1.6-2.7 GHz, dual-core, 3 MB cache, 15 W |
| RAM | 16 GB RAM DDR3L @ 1600 MHz |
| Storage | 240 GB M.2 SSD |
| Operating System | Windows 7 Ultimate (64-bit) |
| Dimensions | 115 mm × 111 mm × 32 mm |
| Weight | ~400 g |

TABLE II
RF FRONT-END TECHNICAL SPECIFICATIONS

| | |
|---|---|
| RF Front-End | Ettus USRP B200 |
| RF Coverage | 70 MHz to 6 GHz |
| Bandwidth | 200 KHz to 56 MHz |
| ADC Resolution | 12-bit |
| Oscillator | GPSDO (OCXO), frequency stability: ±25 ppb |
| Interface | SuperSpeed USB 3.0 |
| Dimensions | 97 mm × 155 mm × 15 mm (board only) |
| Weight | 350 g (board only) |

Table II shows specifications for the front-end. Both front-end and host PC of choice weight around 400-500 g each, thus achieving small dimension and weight. Common internal RF components of the front-end are a low-noise amplifier (LNA), local oscillator (LO), low-pass filter (LPF), and an ADC. Specifications for these components can be seen here [18], [31].

Budget SDR front-ends (including USRP B200) typically come with an internal temperature-controlled crystal oscillator (TCXO), which has limited frequency accuracy. For GNSS satellite signals, a low frequency accuracy oscillator can introduce uncertainties due to phase discontinuities caused by imperfect oscillator frequency accuracy, therefore affecting position accuracy. These phenomena are called cycle-slips [32]. The observed phenomena can cause abrupt channel loss of lock from satellites. To avoid these phase jumps and for better calibration on the front-end oscillator, many SDR solutions in literature propose a 10 MHz reference signal which is used as an external frequency source to replace or aid the local oscillator on the front-end, which is not convenient for mobility. Other SDR solutions propose advanced frequency-disciplined systems based on neural model predictive filters to compensate the local oscillator [33] which aim for ultra-high accuracy on timing applications but require additional software and hardware that lacks portability. Instead, for our approach, a relatively low-cost and accurate oven-controlled crystal oscillator (OCXO) module from Ettus [34], is added to the front-end as a compact board-mounted kit, compatible with the B200 board. This is also called a GPS-disciplined oscillator (GPSDO) clock due to an additional built-in GPS unit that can meet even higher and more accurate frequency and stability requirements than the OCXO by itself if needed due to synchronization with GNSS signals. The proposed GPSDO can lock to satellites in 1 minute and provide stability to the OCXO. The GPSDO mechanism works like a phase-lock loop (PLL) by compensating phase and frequency changes in the local oscillator with respect to GPS satellite signals, as well as adapting to environmental parameters such as temperature, and aging [35]. This board-mounted kit OCXO replaces the internal TCXO on the B200 board, therefore three distinct types of time references are available for experimentation: TCXO (built-in), OCXO (board-mounted kit), and OCXO with disciplining (OCXO + GPSDO). Experimental testing is conducted in a later section to assess the board-mounted kit precision with and without disciplining.

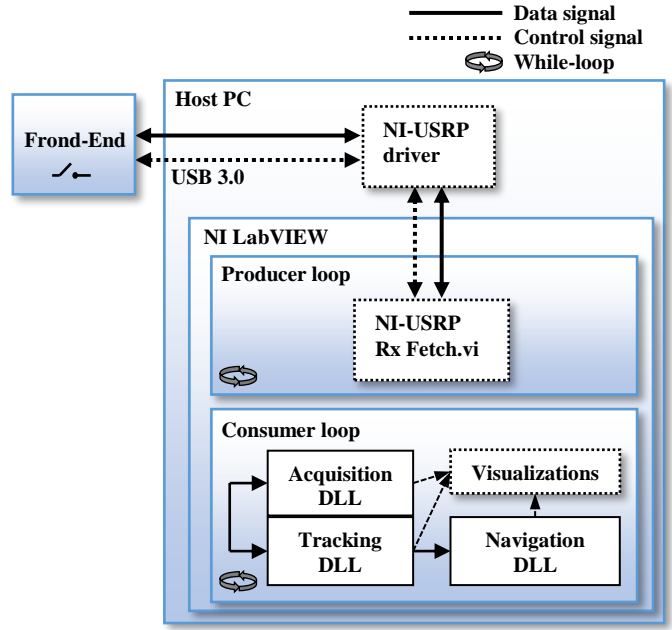

Fig. 3. Front-end and LabVIEW interface via NI-USRP driver. Samples delivered in chunks to DLL baseband modules for post-processing.

### B. Host PC software architecture

The SDR receiver is shown in Fig. 3. The outermost layer of the software part in the host PC is LV-based, which acts as a data flow handling and visualization environment, interfacing the front-end with the internal C/C++ DLL baseband processing modules. The main functionality of the LV interface is in collecting raw digitized samples from the front-end and processing them in real-time to find PVT solution via the baseband modules. The contribution of this paper is demonstrating and characterizing various acceleration techniques for real-time operation in SDR mode for cost-effective research platform purposes. Most of the implemented GPS algorithms are similar to common receiver implementations such as in [2] and [16].

*1) LabVIEW development environment*

LV is a software development environment based on programs called virtual instruments (VIs) which can be visually programmed independently and attached to work either as main or sub VIs. There is always a main VI where all the upper layer functionality is held such as data flow and main execution of loops. Based on this visual programming, LV becomes effective for fast prototyping and data flow handling, thus ideal for SDR solutions. LV has a front panel, which acts as a graphical user interface (GUI) editor where built-in visualizations and controls are available, and a block diagram, where actual visual flow programming is done by wiring built-in and customized function blocks (sub VIs) and loops to achieve visual programming requirements. The flow programming occurs logically from left to right, thus allowing parallelism. LV offers several strengths such as fast prototyping and front-end interface support, but it also presents weaknesses such as reliance to LabVIEW only platform which constrains full-access optimization to some extent, lack of access to its kernel for highly specific optimizations, and its design lacking high-end commercial solutions.

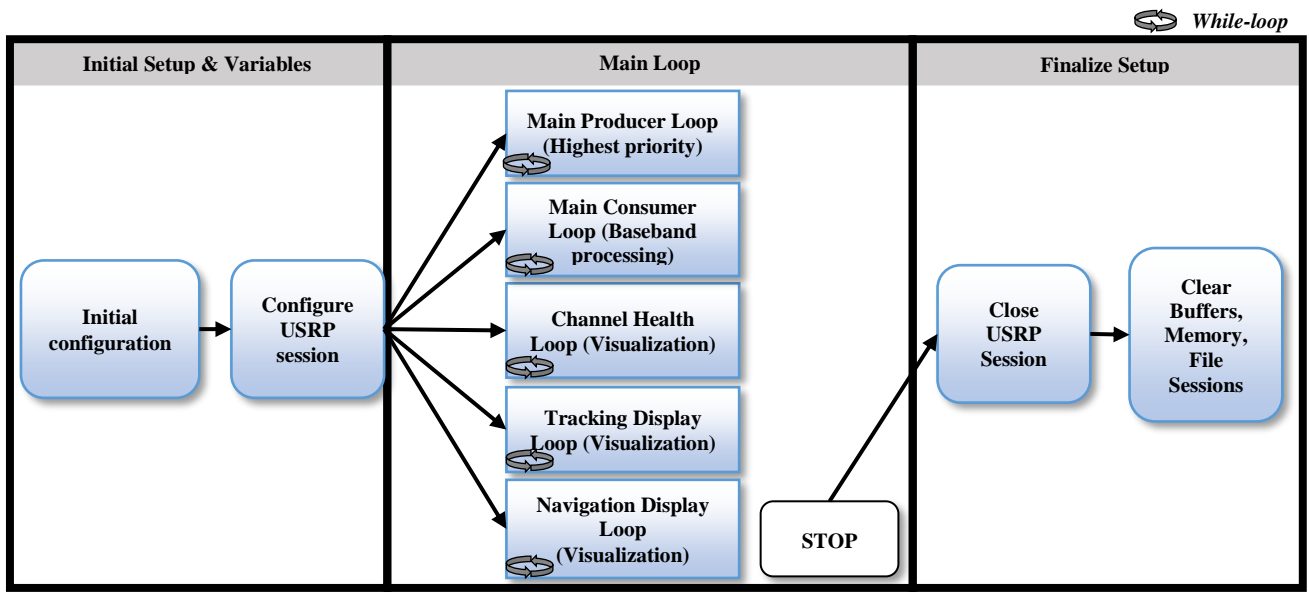

Fig. 4. SDR main VI block diagram architecture consisting of three main stages: initial setup and variables, main loop, and finalize setup. The main loop runs infinitely and consists of *while-loops* running in parallel for different processes, until user halts receiver operation.

LV interfaces with native (NI-USRP) drivers to communicate with the front-end. As seen in Fig. 3, this driver sends control signals to the front-end and fetches data in a pre-configured size of blocks (or chunks) and sends them to the main LV VI, which eventually passes these samples to the DLL baseband modules. This interface is easily handled by LV with built-in NI-USRP configuration and sample fetching VI blocks which configure the front-end and requests blocks of samples. The block NI-USRP Rx Fetch.vi fetches samples from the front-end on every iteration (see Fig. 3).

*2) Main Producer/Consumer loop*

The way LV controls the data flow between the front-end interface and the baseband processing is by a common application design architecture called the producer-consumer loop (see Fig. 3). Based on this design pattern, LV can handle a real-time continuous operation by acquiring data on the producer loop in a non-restrictive and high priority way and sending it to a data queue which allows the raw data to be collected in memory as it is acquired. Then the consumer loop dequeues the data from memory in a first-input-first-output (FIFO) order and sends it to the baseband blocks, i.e. DLL modules, for further processing. Both producer and consumer loops are a *while-loop* structure each running indefinitely and in parallel. These loops operate continuously until the user halts the receiver execution.

The producer loop should be the highest priority of the loop-pair, since it interfaces with the front-end by commanding it to collect raw data chunks (NI-USRP Rx Fetch.vi) in real-time; this, to avoid discontinuities on the incoming GPS signal. Also, if there would be more tasks involved in each loop iteration of the producer loop, overflows and underflows could easily be triggered since the front-end is expected to collect data without interruptions and typically lacks an internal buffer. The internal queue or buffer utilized between this loop-pair is automatically handled by LV in terms of memory allocation, thus occurring in the background. The consumer loop should also be capable of processing incoming data in real-time so that it outputs up-to-date user location.

*3) DLL integration*

Conventional GPS baseband functionality can be divided into three main modules: acquisition, tracking, and navigation. Each GPS baseband module includes numerous relevant C/C++ functions which are compiled into DLLs for further integration in LV environment. Full optimization is done at compilation by, e.g., Visual Studio C/C++ 64-bit compiler. Once generated, these DLLs can be accessed by LV's built-in *Call library function node*. When called upon, LV sends relevant input arguments to these functions such as clusters of data types, which are compatible with C/C++ data type structures. The input arguments can be passed by reference, therefore omitting duplicates of the input data structures on every function call.

*4) Main VI block diagram data flow*

Fig. 4. illustrates the contents of the main VI block diagram. The flow inside the main VI consists of three stages: Initialization of data structures and variables, the main loop (multi-producer/consumer loop), and the finalize setup and close session. The initialization part oversees allocating and initializing memory on two main data structures (clusters in LV) which act as pipe flows that connect to relevant function blocks as input arguments: the system configuration which is a structure of shared global configuration parameters relevant to GPS L1 parameters and current receiver session (e.g., sampling rate, receiver gain, etc.), and the channel structure containing specific variables linked to a locked channel. Each channel structure contains a structure of variables which keeps track of the channel status, health, tracking loops, Doppler, Ephemeris and other essential information associated to a satellite lock. Eventually, all the channels are gathered into an array and therefore flow in a single line or wire. After these two main variable pipe flows are initialized, the USRP session is configured. This involves communicating to the front-end via the USB port to properly initialize and setup the receiver gain, sampling rate, data block fetch size, data format (e.g., INT8),

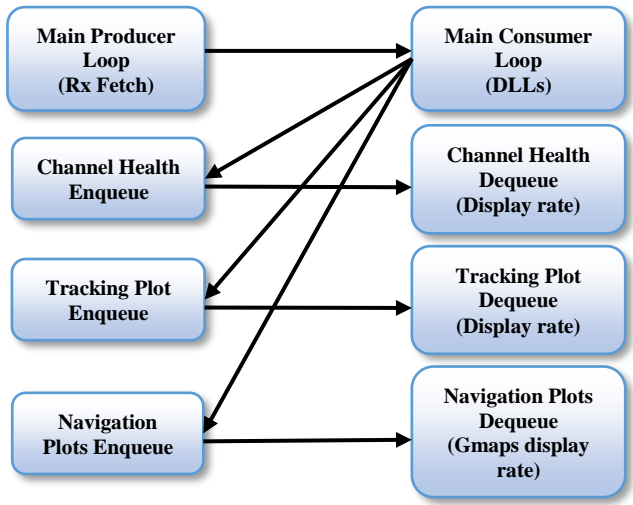

Fig. 5. Multi-producer/consumer loops interaction for multi-threading and independent data flow rates.

reference clock, and other relevant settings for current session. Once initial configuration is done, the main producer/consumer loop is where the receiver operates in real-time until the stop button is toggled. When stopped, the VI goes to the finalize setup stage where it closes current USRP session, frees previously allocated memory on buffers and structures, and closes input and output file sessions for log files if configured.

*5) Multi-producer/consumer loop for multi-threading*

As seen in Fig. 4, multiple producer and consumer loop-pairs are being used. A multi-producer/consumer loop mechanism was added to increase multithreading which relieves overhead computational loads for concurrent tasks such as real-time visualizations. Now independent rate configuration is possible for each visualization. Multi-threading is accomplished by creating several lower priority producer/consumer loops generated in the main consumer loop, therefore acting as the producer. Otherwise the main consumer loop should process the incoming GPS signal and generate output visualizations sequentially on each iteration. This sequential processing affected every iteration of consumer loop processing time and ultimately created bottle-neck effects. With multi-producer/consumer loops added, the main consumer loop is now responsible to send (produce) output data to other three loop-pair queues: the tracking display loop, the channel health loop and the navigation display loop, and its tasks are therefore lightened. Then, these low-priority consumer loops can now post-process, or in this case, display output visualizations at an independent rate which can be configured.

Fig. 5. shows the interaction of these loop-pairs. Each loop-pair is a thread in LV and can have a different priority and rate. Also, each loop-pair is assigned a buffer that inputs data blocks into memory (producer) and then process this data (consumer) at different rates. Therefore, these rates can be independently configured based on user requirements. LV uses built-in Queue/Dequeue function blocks that are used for sending data to buffers automatically allocated to each producer/consumer loop-pair so independent tasks can handle data from these buffers in an automatic multi-threading setting. Therefore, inside the main consumer loop, there are three Queue blocks (three producer loops inside the main consumer loop) that are sending data to each of these three categories, but the visualization does not execute until these lower priority consumer loops are activated. These loops are activated with a delay timer to control the rate.

*6) Main consumer loop and baseband modules*

Fig. 6 shows a detailed view on the main consumer loop where all baseband processing occurs. This is the core of the SDR GPS receiver where most computation processing occurs. This flow occurs on every *while-loop* iteration, and quasi-sequentially because of LV's multi-threading and parallelism. The Dequeue block collects a fixed-size of samples from the first-input-first-output (FIFO) buffer coming from the producer loop and sends it to the acquisition and tracking blocks. The two main flow pipes: system configuration and channel parameters are updated accordingly on each loop iteration by setting them as shift registers (SR) as seen in Fig. 6. This means, once one iteration is done, both pipe flows are recirculated to the initial

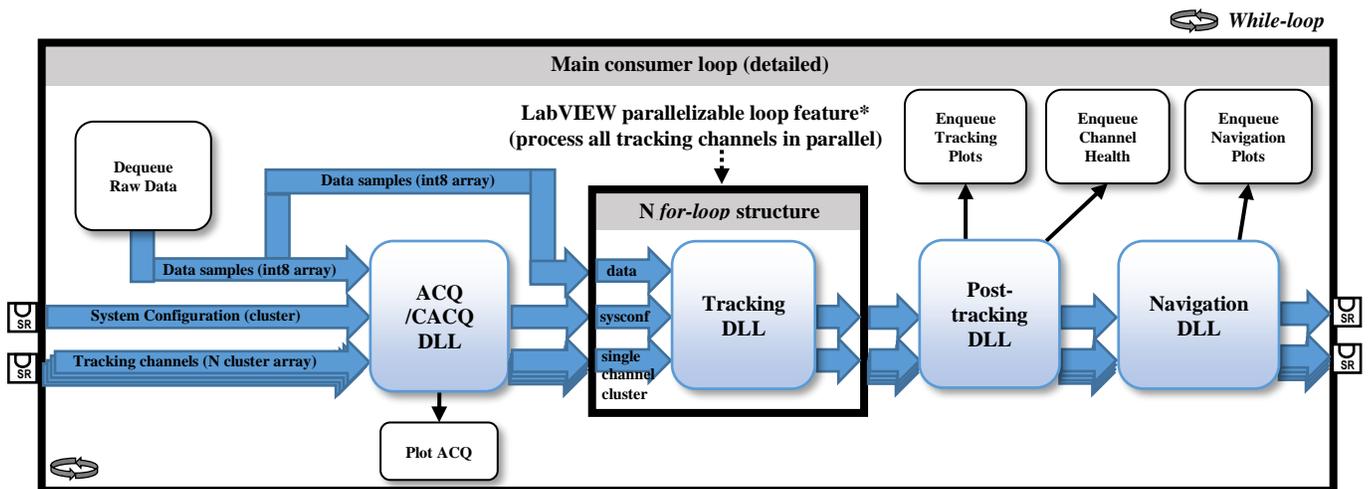

* LabVIEW parallelizable loop feature allows concurrent execution of DLL library on each channel if channel cluster variables are independent to each other (see Fig. 7).

Fig. 6. Main consumer loop detailed showing all three main pipe flows: data samples, system configuration, and channels, passing through baseband modules.

left side of the structure retaining modified value output from the block modules, ready for next iteration.

As for the baseband modules, first the *Acquisition* module checks whether the receiver is lost (i.e. no channels are being tracked) and/or if a set of quality metrics are held, for it to run fast acquisition algorithms. The acquisition/continuous acquisition (ACQ/CACQ) block runs through a set of quality-of-service (QoS) metrics when deciding to run acquisition algorithms, all of them having same priority: (1) number of tracking channels is less than 4 (minimum required for PVT solution); (2) percentage number of tracking channels drop rate (for instance, if 30% of initial number of acquired channels drops in a short period of time, denoting a blockage or so); and (3) at least <min_num_chans> with <min_cn0> (i.e. at least 5 channels with 40 dB-Hz). If acquisition is toggled to run (which is always the case at initial receiver execution), it then first requires filling its buffer depending on the coherent integration length setting (i.e. 4 ms of data), before algorithm execution. The *N* found satellites are assigned to one out of 12 total tracking channels and the system configuration modifies the state to *Tracking*. If continuous acquisition is triggered later, the system returns to ACQ/CACQ block for a parallel acquisition not affecting current tracking channels and assigns new channels accordingly (see Fig. 6). When ACQ/CACQ algorithm is executed, visualization table is updated with results.

For *Tracking* and *Navigation*, on each consumer loop iteration, data sample chunks (an array of raw samples) obtained from the data queue are sent in parallel to the tracking baseband module (as well as acquisition). There should be *N* active tracking channels processed with the current data samples. This is the most computationally intense part of the SDR receiver as it should process data in real-time since GPS signals are continuously being received and a missed data sample chunk could signify a loss of lock in most channels. The post-tracking block collects relevant data from all channels and translates it into channel health statistics, as well as real-time tracking plots for Doppler frequency in Hz, navigation bits vs time, and in-phase vs. quadrature (I-vs-Q) graph also known as constellation diagram. Finally, navigation module extracts relevant information from all *N* tracking channels and computes a PVT solution if available. Position solution is enqueued and should be displayed in a latitude vs longitude graph, as well as an integrated Google Maps visualization which uses a web Google Maps API. For Google Maps display, an active internet connection is required in the host PC to download display actual maps from Google servers.

*7) LabVIEW and C/C++ memory compatibility*

Since there is an interchange of variables, clusters, and other data structures between two programming environments, i.e. LV and C/C++, when exchanging data between both environments via DLLs which are typically passed as reference, three concepts should be considered: 1) memory alignment, 2) data type compatibility, and 3) order of data types in a structure.

Memory alignment between data types should be fully compatible to avoid fatal errors such as data corruption and memory access violations. LV 64-bit and Visual C++ 64-bit are both compatible with same memory alignment method called "natural alignment". This type of memory alignment takes the biggest data type in the structure, typically 8 bytes (64 bit) for a complex data type, and uses it as a base to line up with other data types. This means, all data types smaller than 8 bytes (i.e. *char* data type which is 1 byte) will still take 8 bytes of space in memory by adding padding or dummy bytes, thus making each memory address a multiple of 8 bytes, for a given data structure.

TABLE III
LABVIEW AND C/C++ DATA TYPES COMPATIBILITY

| C/C++ | LabVIEW | Size in bytes |
|---|---|---|
| char | I8 | 8 |
| uchar | U8 | 8 |
| char * (64-bit pointer) | I64 (64-bit pointer) | 64 |
| int | I32 | 32 |
| unsigned int | U32 | 32 |
| double | DBL | 64 |
| char[4] | Cluster of four I8 | 32 |
| struct | Cluster | variable |

Another important consideration is the data types used interchangeably between both programming environments. LV has its own data type names that are compatible with C/C++ data types. Table III shows data type compatibility between both environments.

Finally, the order of the variables inside a generated data structure is also relevant between both platforms, as they should match on both environments to avoid data corruption. As an example, if a cluster (struct in C/C++) is created in LV with two *I8* and then a *DBL* variable types, this same order should be reflected in C/C++ as two *char* variables, then a *double* variable sequentially.

*8) LabVIEW parallelizable loops*

*Parallelizable loops* are a LV feature that is similar to the inherent parallel scheduling mentioned before, but should be assigned manually onto *for-loop* structures. Parallelizable loop candidates can be assessed by LV based on the block diagrams and dependency between input and output variables, but ultimately are assigned manually by the developer. A very useful parallelization loop is in the tracking channels, since each satellite channel can run independently and accomplish speed gains to compete for real-time operation. There are ways to parallelize code in C/C++ coding but requires high programming skills to do so, as opposed to LV's Parallelizable Loops feature which can be assessed and configured within LV's block diagram. There should be no resource dependency between each tracking channel, when applying this feature to avoid fatal programming errors such as memory access violations.

Fig. 7 shows the adjustments to the receiver to incorporate the Parallelizable Loops accelerator feature. Initially, the baseband tracking DLL call function was processing all tracking channels by using an internal C/C++ *for-loop* structure, therefore the input and output variables from LV to the DLL were an array of channel structures. The DLL tracking call function code was internally modified so that it would process only one channel structure, thus leaving the *for-loop* structure previously found in the C/C++ code, to LV, as seen in Fig. 7. This LV *for loop* structure takes as input the same array of channel structures with size *N* and extracts one by one to send it to the DLL call function node, then regroups them at the end of the structure. Of course, all channels are now processed concurrently with the Parallelizable Loops feature, therefore

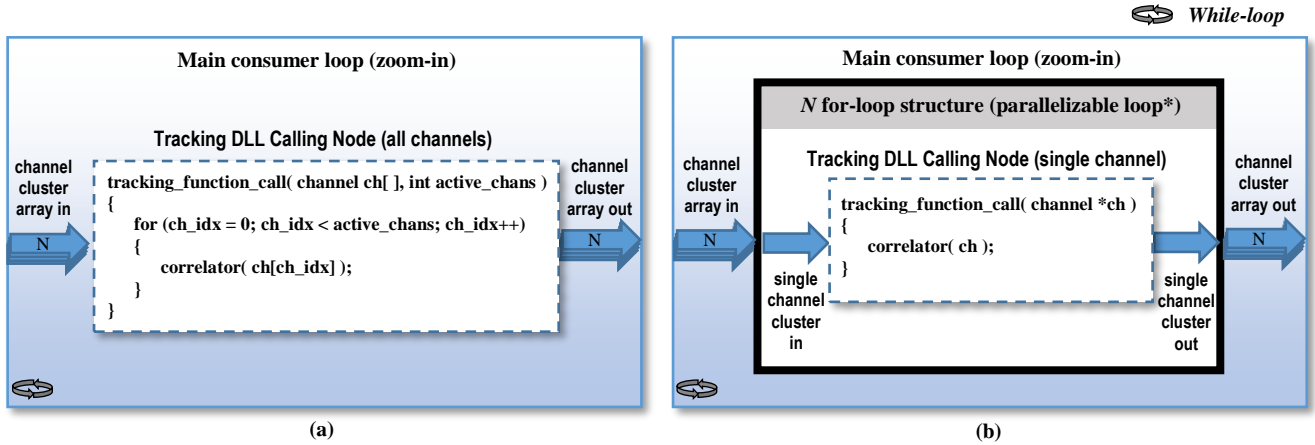

Fig. 7. LabVIEW Parallelizable Loops feature applied to tracking baseband DLL module for individual channel processing and overall computational acceleration. (a) shows sequential implementation of tracking channels on C/C++ *for-loop*, (b) shows Parallelization Loops applied to built-in LabVIEW *for-loop*.

TABLE IV
PROPOSED SDR ACCELERATION FACTORS

| Acceleration factor | Source |
| --- | --- |
| Algorithmic accelerator | Software-based |
| Single instruction, multiple data (SIMD) | C/C++ and Intel processors |
| Optimized libraries | C/C++ libraries |
| Parallel/multi-core scheduling | LabVIEW-based |
| DLL integration | LabVIEW-based |
| Inherited multithreading | LabVIEW-based |
| Data acquisition, data flow control | LabVIEW-based |
| Real-time advanced visualizations | LabVIEW-based |
| FPGA and DSP | Hardware-based |

accelerating each consumer loop iteration when dealing with tracking loops.

## IV. ACCELERATION FACTORS AND CONFIGURABILITY

Table IV shows a summary of acceleration factors that leverage the proposed SDR receiver's computational power for real-time operation. Most of the implemented features are exploited from LV platform built-in features. Many SDRs use a combination of acceleration factors as found in Table IV, so a brief description of the proposed SDR features follows.

*Algorithm accelerators* include the advanced acquisition module [26] implemented in C/C++ as a replacement to PCS algorithms which shows dramatic gains in joint computations.

*Single instruction, multiple data* (SIMD) processing is also implemented in the DLL functions and resulted in dramatic computational gains in the tracking loops. SIMD are specialized, built-in machine language instructions (or opcodes), mostly found on modern Intel 64-bit processors. These assembly-coded functions can be implemented directly by using Intel Intrinsics [36] which are function calls in C++ language. The benefit of SIMD is the direct usage of multi-purpose processors' dedicated registers capable of parallel complex arithmetic operations such as table mapping, dot products, accumulators, and other parallel operations resulting in speed increase. The receiver's latest version v5.0 leverages from SIMD features since its baseband algorithms are developed in C/C++ language.

*Optimized C/C++ libraries* feature matching internal architecture for specific resource utilization based on target host PC. As much as threefold acceleration was achieved on proposed SDR receiver for FFT routines by implementing optimization libraries such as *FFTW* [37] when compared to other libraries such as k*issFFT* [38]. Eigen [39] was another library used which specializes in matrix operations for least-squares (LS) computation relevant to conventional PVT solutions. These optimized libraries have internal functions that try to exploit as much as possible the host PCs architecture and processor capabilities. These libraries can, for instance, apply built-in SIMD functions to the FFT routines, or schedule computations for multi-threading execution based on number of cores in the processor.

LV has characteristic functionalities that effortlessly assign *parallel tasks to multi-core processors* by taking advantage of the visual block diagram programming style. Program parallelism achieves performance gains by concurrently running several independent block diagram paths which occur transparently to the developer since LV's compiler takes care of scheduling these processes. The compiler also recognizes the host PCs' capabilities in terms of cores and threads available. The SDR receiver is strongly tied to LV's visual programming for real-time processing and parallelization due to this feature. In addition, the Parallelizable loops feature was used in the SDR tracking loops as mentioned in previous section.

Based on *DLL integration* into LV, the proposed SDR uses baseband modules that were successfully implemented in C/C++ with their respective data type compatibilities.

LV uses inherent multi-threading which applies also to data acquisition and data flow. All these concepts are integrated when using the previously described producer/consumer loop-pairs. Also, visualizations are included in this integration.

Many real-time SDR GPS receivers use *FPGA* accelerators [21] for common functions such as FFT-based acquisition routines. For simplicity in hardware components, this SDR was chosen to have all functionality in software and exploit host PC architecture as mentioned in previous sections.

TABLE V
PROPOSED SDR CONFIGURATION OPTIONS

| Configuration | Units | Options |
|---|---|---|
| Sampling rate | MHz | 2, 4, 5, 10, 20, 25 |
| Receiver gain | dB | 0-30 |
| Satellites to search | integer | 1-32 |
| No. tracking channels | integer | 1-12 |
| Acquisition search band | KHz | 10, 12, 14, 16, 18, 20 |
| Acquisition coherent integration length | milliseconds | 1, 2, 4, 8, 16, 32 |
| PVT update rate | Hz | 1, 2, 5, 10, 20 |
| PVT averaging depth | samples | 0, 10, 20, 50, 100 |

TABLE VI
LIST OF SDR INSTRUMENT MEASUREMENTS AND VISUALIZATIONS

| Measurement/Visualization | Details |
|---|---|
| Acquisition table | PRN, Doppler frequency, code-phase, detection ratio. |
| Tracking display | Navigation bits v. time chart, constellation diagram, Doppler frequency v. time chart. |
| Channel health display | PRN, carrier-to-noise ratio (CNR), carrier lock ratio, lock failure count, channel state, valid for PVT flag. |
| Navigation chart | Latitude v. longitude chart, GDOP, RMS error (meters). |
| Google Maps | Zoom level, map type. |
| Tracking logging output file | Output .log file for debugging |
| Navigation logging output file | Output .log file for debugging, output .kml file compatible with Google Earth software. |

### A. Software configuration and instrumentation measurement output

The proposed SDR has several configuration options that are compatible with the USRP B200 front-end as well as the host PC capabilities. Table V lists the available configuration options which can be used in online (real-time) operation mode as well as in offline mode with a pre-recorded file. The receiver operates only in INT8 mode, and for offline mode can chose between two types of sample inputs: in-phase (I) only, and in-phase and quadrature (I-Q) interleaved samples. The former is used for when an intermediate frequency (IF) is used in the front-end such as in [40]. This IF can be specified (in Hz) when running in offline mode, and all visualizations can be utilized the same was as in online mode. Since the proposed front-end (USRP B200) uses a direct down-conversion (DDC) system, there is no IF involved and captured samples are I-Q interleaved as follows: $s_0^I, s_0^Q, s_1^I, s_1^Q, s_2^I, s_2^Q, \ldots$, which are already in baseband level. The receiver has a maximum of 12 tracking channels, however this could be expanded in future releases. The *PVT update rate* defines the number of position samples outputted per second, and the *PVT averaging depth* uses an averaging sliding window on PVT output samples, thus smoothening final user position. The rest of the configurations are conventional on a GPS receiver.

Table VI shows common GPS receiver instrumentation measurement outputs as well as visualization outputs included in the proposed SDR. Most outputs are well-known measurements for real-time visualization and monitoring of GPS health channels, as well as other statistics. Measurements such as carrier-to-noise ratio (CNR), carrier-lock ratio, and channel state, are found in the *channel health display* and their refresh rate for each visualization and/or instrument can be independently configured. Logging outputs for tracking channels and navigation outputs such as *.kml* file for Google Earth can be configured for output when session finalizes.

TABLE VII
SUMMARY LIST OF SDR VERSIONS BASED ON CHARACTERISTIC ACCELERATION FACTORS

| Version | Characteristic acceleration factor or feature |
|---|---|
| First version | Google Maps |
| Second version | Multiple-producer/consumer loops |
| Third version | LabVIEW Parallelizable loops |
| Fourth version | SIMD feature |

## V. COMPARATIVE RESULTS

This chapter discusses several comparative results of the proposed SDR against other open-source receivers such as *fast-gps* [16], and *GNSS-SDR* [25], to assess overall performance of the receiver. It also compares against several acceleration factors of the receiver. Table VII summarizes the most important acceleration factors taken into account for comparison results, which differentiate cumulative upgrades between proposed receiver concepts (i.e. each newer version contains previous features). The results will be roughly divided into the three baseband modules: acquisition, tracking, and navigation. In most cases, the receiver was used in the following configuration parameters: sampling rate of 5 MHz, acquisition coherent integration of 4 milliseconds, acquisition search radius of 10 KHz, among other settings. Also, since DLL functions are C/C++ based, certain tests were made in command line interface (CLI) only, as well as in the complete LV-based receiver, since no front-end interface is required for offline test and benchmarking can be computed for certain tracking loops and acquisition algorithms. This final testing scenario (offline mode) is desired to assess possible LV overhead when compared to CLI only receiver (DLLs and executable).

### A. Open-source GPS receiver alternatives

The open-source receiver, *GNSS-SDR*, was selected for comparative results against the proposed SDR. The *GNSS-SDR* version used for this open-source receiver was *0.0.6*. *GNSS-SDR* works in Linux environment and it is heavily dependent on GNU-radio framework [17] as well as other dependencies for installation. This receiver alternative was chosen since it's one of few real-time open-source receivers available for online comparisons against the proposed SDR. Although *GNSS-SDR* is a CLI only therefore lacking advanced real-time visualization aspects, it allows for output logging of several channel parameters. *GNSS-SDR* is also compatible with same host PC and front-end (NUC and B200), therefore, same computational resources, sampling rates, receiver gain, and other similar parameters can be used as a direct comparison. Several tests to assess performance, robustness, precision, among others are desired. The other alternative, *fast-gps*, was chosen and used as a reference receiver for the development of the proposed SDR. Since *fast-gps* works only in offline mode, this receiver was used on similar comparisons for benchmarking and other features.

TABLE VIII
PERFORMANCE ACQUISITION IMPLEMENTATION ON THE HOST PC WITH
LV-BASED RECEIVER

| Acquisition timing (msec) | PCS acquisition (4 msec) | Advanced acquisition (4 msec) | Advanced acquisition (8 msec) |
|---|---|---|---|
| LV-based blocks | 27,544.3 | 504.1 | 1,038.8 |
| *kissFFT* (DLL function) | 7,982.6 | 331.4 | 674.4 |
| *FFTW* (DLL function) | 3,493.6 | 166.3 | 348.3 |

TABLE IX
PERFORMANCE ACQUISITION IMPLEMENTATION ON THE HOST PC WITH
CLI ONLY RECEIVER

| C++ CLI acquisition timing (msec) | PCS acquisition (4 msec) | Advanced acquisition (4 msec) | Advanced acquisition (8 msec) |
|---|---|---|---|
| *kissFFT* | 7,850.8 | 347.2 | 692.7 |
| *FFTW* | 3,466.8 | 181.9 | 355.5 |

*B. Acquisition comparative results*

Significant comparisons in acquisition are made in this section. One can summarize them into four categories: (1) platform performance, (2) optimized libraries performance, (3) advanced algorithm performance, and finally (4) LV overhead performance. All said comparison dimensions are integrated into acquisition comparison tables.

With respect to platform performance, acquisition algorithms are developed in C/C++ compiled and optimized versions as well as with LV-based blocks such as FFTs. This to assess comparisons in timing when algorithms are built and compiled with either platform. For optimized libraries comparisons, *kissFFT* [38] and *FFTW* [37] libraries are compared as a second dimension. For a third dimension, reference receiver *fast-gps* uses a conventional PCS acquisition algorithm, and since an advanced acquisition [26] is used in place, a comparison is made in this respect. Finally, for a possible LV overhead performance, a C/C++ only (CLI) version is ran to compare optimized libraries and algorithms. For all tests, 4 milliseconds of integration length were used, as well as 10 KHz acquisition search band. Also, an offline recording file was used with 12 visible satellites. Only the acquisition algorithm along with data fetching was benchmarked for a fair comparison, thus avoiding variable initializations, and other steps in the programs. Data fetching was included specially for LV overhead comparisons.

Table VIII shows comparisons between platforms, algorithms, and optimized libraries. With respect to platform, C/C++ optimized functions can be as much as 7.8 times faster than built-in LV blocks when using fastest optimization library. Also, between the optimized libraries *kissFFT* and *FFTW*, the latter is more than twice as fast as the former, for all cases. For the advanced algorithms, as much as 54 times faster can be seen when comparing algorithms developed in LV-based blocks, and around 22 times faster when C/C++ DLLs are used in LV receiver. Finally, an 8 msec acquisition integration length was compared, which show two-fold slower in timing for all dimensions (this is expected) when compared against 4 msec integration length, but at the same time, still maintaining drastic improvements of up to 26 times faster than PCS algorithm with 4 msec, for LV-based algorithms, and up to 11 times faster with C/C++ DLLs.

Table IX aims to compare acquisition algorithms, optimized libraries, and at the same time, compare against Table VIII for possible LV overheads, since tests in Table IX were made in CLI only. In average, numbers seem very similar, thus showing little to no computation overhead when using LV with DLL integration against CLI only offline receiver for acquisition algorithms.

*C. Tracking comparative results*

Since tracking is considered the highest computational cost and most time-critical operation for GPS receivers, relevant comparisons are made in this section. Important comparisons were divided into two categories: online and offline tests. Tests aim to assess proposed receiver performance, robustness, precision, real-time operation, among others. For offline testing, three important comparisons were made to demonstrate robustness and configurability of the receiver. For online mode, two comparisons were made to assess CPU load, memory, and other metrics. In total, five comparison tests were evaluated.

*1) OFFLINE: Local replica carrier wave configurability*

Tracking correlators continuously generate local replicas for carrier and code phases for synchronization, based on discriminator output parameters. Since these local replicas should be generated continuously and in real-time, computation efforts for host PC can be high. Typically, quantization of carrier waves stored in generated look-up tables (LUTs), are used for faster computations, while at the same time sacrificing channel quality. In addition, the proposed SDR tracking correlators not only use integer arithmetic (INT8 for input samples) for faster computations, but also can choose different carrier wave generation methods. Comparative results between several carrier wave generation methods, beginning from conventional C/C++ floating-point *sin()* function, are evaluated. The second option is a floating-point exponential series approximation of the *sin()* function, which uses only the first two terms of the sine and cosine Taylor expansions. The next three options are related to 16, 8, and 2 value LUTs.

For carrier wave quantization tests, the tracking integration lengths was 1 msec (epoch). Relative loss in decibels (dB) metric compares against floating-point *sin()* function to assess performance and robustness. An offline recording (same as used in acquisition tests section) was used, running 300 seconds and averaging dB loss for 12 channels when comparing relative loss in carrier wave generation methods and timing to obtain an experimental evaluation of the complexity of computations for each method.

Table X shows results for carrier wave configurability in proposed receiver in offline mode. Performance times are in nanoseconds. The average time per epoch is 1 msec, and it is normalized per channel, since recording file showed 12 tracking channels during all times. The number of real-time tracking channels shows tracking-only complexity of computations, but lacks a possible online overhead such as LV front-end interface, data acquisition, among others. The number of real-time channels was found by dividing average time per epoch per channel to a single tracking integration length (1 msec), which would show real-time operation. The exponential series

TABLE X
OFFLINE REPLICA CARRIER WAVE QUANTIZATION PERFORMANCE ON HOST PC

| SDR Tracking Quantized Carrier Loss (dB) | dB Loss (relative to base) * | Avg. time per epoch per channel (nsec) | No. real-time tracking channels* |
|---|---|---|---|
| sin() function (base) | 0.00 | 167,906.67 | 5.96 |
| Exponential series approximation | 0.00 | 87,833.67 | 11.39 |
| 16 levels LUT | 0.05 | 76,105.12 | 13.14 |
| 8 levels LUT | 0.53 | 75,920.97 | 13.17 |
| 2 levels LUT | 1.15 | 67,968.07 | 14.71 |

* Always at 5 MHz sampling rate (NUC + B200 + NI simulator)
* dB Loss relative to built-in C++ floating-point sin() function

TABLE XI
OFFLINE TRACKING EPOCH PERFORMANCE ON HOST PC FOR DIFFERENT ACCELERATION FACTORS

| SDR Tracking Time (nsec) | Avg. time per epoch per channel (nsec) | No. of real-time tracking channels* |
|---|---|---|
| Google Maps version | 75,920.97 | 13.17 |
| Multiple-producer/consumer loop feature | 75,787.05 | 13.19 |
| Parallelizable loops feature | 40,258.45 | 24.84 |
| SIMD feature | 11,125.70 | 89.88 |

* Normalized number of tracking channels by dividing by total amount per tracking channel by an epoch's time, i.e. 1 millisecond.

approximation is 1.91 faster and barely sees any relative loss when compared to base sin() function. The 16 level LUT shows the best performance gain (2.2 times faster) while at the same time keeping relative loss to a minimum (0.05 dB). Still, for all tracking tests, the 8 level LUT was used for simplified debugging purposes, among others.

*2) OFFLINE: Acceleration factors on tracking loops*

As seen in Table VII, four acceleration factors were summarized for comparison purposes. The acceleration factors test aims to see performance gains in timing when comparing these acceleration factors for different cumulative versions of the receiver, i.e. each newer version contains previous features. Using similar comparison metrics as seen in Table X, tests were performed with 8 level LUT, 5 MHz sampling rate, and same offline recording file with 12 channels.

Table XI shows results for cumulative receiver acceleration factors. Between first and second version, a small gain is seen since the multi-producer/consumer loop factor aims to increase multi-threading and allow more visualizations in LV receiver, and not to accelerate actual tracking loops. Thus, similar performance is seen for both versions. For the Parallelizable loops feature, the average time per epoch per channel metric is 1.88 faster than first version, therefore almost doubling the total number of real-time channels that the receiver can support in real-time operation at present configuration and host PC. Finally, the SIMD feature version performs 6.82 faster than first version, thus dramatically increasing the number of possible real-time tracking channels to almost 90 when using 5 MHz sampling rate.

TABLE XII
OFFLINE OVERALL RECEIVER PERFORMANCE COMPARISON ON HOST PC FOR SAME LENGTH RECORDING FILE

| Process OFFLINE file 300s recording. This file is 12 channels all 300 seconds | Time (sec) |
|---|---|
| Proposed SDR, LabVIEW-based | 51.292 |
| Proposed SDR, CLI only | 49.868 |
| GNSS-SDR | 310.843 |
| fast-gps | 179.302 |

*3) OFFLINE: Overall receiver performance benchmarks*

This test compares overall receiver benchmark from the time of execution to finalization for several receivers in offline mode: the proposed receiver in LV with DLL integration, proposed receiver in CLI only, GNSS-SDR, and fast-gps. This way, all variable initializations, computations, and other algorithms are measured in performance timings. For the test, the same offline recording file was used, which broadcasts 12 satellites during recording, thus simulating a 5 minutes signal. Also, the proposed SDR uses the latest acceleration features and all tests were performed in host PC.

Table XII shows results for overall receiver performance. For offline mode, LV adds a small overhead of 2.9% increase in time when comparing with proposed receiver in CLI only mode. At the same time, both proposed SDR modalities outperform fast-gps and GNSS-SDR receivers, running 3.5 and 6 times faster, respectively.

*4) ONLINE: Overall receiver performance metrics*

An overall assessment of receiver robustness and performance was measured in the following tests in online mode of operation. The testing measurements were for CPU load percentage, memory occupancy, number of threads, number of real-time channels in a stable operating point at 5 MHz, and maximum number of channels on maximum sampling frequency.

For the number of channels, the aim is to find a stability point in the receiver where it operates in real-time with no crucial delays, overflows, or lost packets. Although 12 channels are generated from NI GPS simulator, both receivers can configure and limit the total number of actual tracking channels. The tests were compared against latest version of proposed receiver and GNSS-SDR, which also has online operating capabilities. For the proposed SDR, a careful computation of whether producer and consumer loops were operating at similar pace, was assessed to decide how many real-time channels the receiver was capable to handle. For GNSS-SDR, numerous tests were assessed until a more-less stable point of operation was observed, with minimal overflow occurrences: this stability point differentiated between uncontrollable overflows rendering the receiver non-operational, and a more-less continuous operation of GNSS-SDR. For the maximum number of channels at a given sampling frequency, no limit was set, and the best performance was assessed for both number of channels and sampling frequency (having 12 channels as a limit for proposed SDR).

For the online tests, a total of 10 executions of 5 minutes each (total of 50 minutes of operation) was evaluated. Similar configuration parameters were used throughout the tests such as: NI GPS simulator broadcasting 12 satellites signal, 5 MHz sampling frequency, similar receiver gain, same RF hardware and antenna, 4 msec acquisition coherent integration lengths, 10 KHz acquisition search band, among other parameters.

TABLE XIII
ONLINE OVERALL RECEIVER PERFORMANCE COMPARISON ON HOST PC FOR SEVERAL METRICS

| Online testing at 5 MHz | CPU load (%) | Memory (KB) | No. of threads | No. of channels in real-time @ 5 MHz | Max. no. of channels in real-time @ Fs |
|---|---|---|---|---|---|
| Google Maps version | 15.99 | 167,864.88 | 34 | 11 | 11 @ 5 MHz |
| Multiple-prod./cons. loop feature | 15.52 | 155,346.13 | 39 | 11 | 12 @ 4 MHz |
| Parallelizable loops feature | 22.73 | 155,248.85 | 39 | 12 | 9 @ 10 MHz |
| SIMD feature | 9.85 | 155,613.47 | 39 | 12 | 8 @ 25 MHz |
| *GNSS-SDR* | 42.98 | 130,529.47 | 64 | 5 | 5 @ 5 MHz** |

\* Online testing using NUC + B200 + NI simulator.
\* Online testing performed at 5 MHz sampling rate for all receivers, as well as 4 msec. coherent integration length for acquisition.
\*\* Maximum number of channels with minimal overflow occurrences and stable operation.

Table XIII shows comparative performance results for online operation for multiple acceleration factors, as well as *GNSS-SDR* receiver.

From CPU load perspective, proposed SDR began gaining load as versions increased, but on the last version with SIMD feature the load decreased by more than two-fold. This is because tracking correlator arithmetic operations are now handled by internal SIMD registers on host PC, making multiple operations at a time while consuming less CPU resources with high efficiency. For *GNSS-SDR*, a quick glance at the high load is shown due to floating-point operations in acquisition and tracking algorithms, as well as many configuration options and a strict dependency on GNU-Radio and other numerous dependencies which require a high number of threads to be instantiated when executed.

From memory occupancy, all receiver showed a similar performance which is minimal compared to host PC's total memory of 16 GB. On threads' perspective, proposed SDR gained 5 threads since the multi-producer/consumer loop feature version. *GNSS-SDR* showed higher thread occupancy, again, due to many instantiations when executed.

At 5 MHz sampling rate, proposed SDR gained tracking channel capacity as versions increased. If comparing online tracking channel capacity against Table XI, one can analyze an overhead cost from LV environment which includes data acquisition, USRP interfacing, visualizations, among other reasons. For the first and second versions, two real-time tracking channels are traded for several LV-based receiver benefits. For the maximum number of channels, since the proposed receiver's limit is 12 channels, sampling rate was used as a variable for finding the maximum optimal operating point of all versions. For the latest version of proposed SDR, a total

TABLE XIV
ONLINE RECEIVER OVERFLOW ROBUSTNESS COMPARISON ON HOST PC

| Overflows (OV) in 50min run (real-time) | No. of OVs | Avg. OVs/sec | Avg. time between OVs (sec) |
|---|---|---|---|
| Proposed SDR | 0.00 | 0.00 | 0.00 |
| *GNSS-SDR* | 40.00 | 0.01 | 70.10 |

of 8 channels at 25 MHz were able to operate in real-time for selected host PC and hardware.

*5) ONLINE: Data interruptions and overflows statistics*

For this online test, latest version of proposed SDR and *GNSS-SDR* were tested for overflows, with same parameters as previous test. A total of 10 runs of 5 minutes each (totaling 50 minutes) in real-time operation, with NI GPS simulator and 12 satellites. Using results from Table XIII to obtain stability point on both receivers, overflows and data interruptions were evaluated. Table XIV shows statistics for data interruptions on both receivers. The proposed SDR showed no overflows. *GNSS-SDR* showed 40 overflows on 50 minutes of operation, which averages to 0.01 overflows per second. Another statistic measured was the average time between overflows, which was found to be 70 seconds. This means, when running receiver, in average every 70 seconds it will overflow and channels will be lost.

*D. Navigation comparative results*

For GPS receivers, navigation precision is an important statistic that characterizes receiver performance. Two statistics were evaluated for precision: root-mean square (RMS) error (in meters), and true mean error. PVT solutions usually output user position in several coordinate systems. The x, y, and z cartesian coordinates for user location were used to measure both statistics. RMS error in this evaluation corresponds to the standard deviation, or how much position measurement samples change between one another. For the true mean error, since the NI GPS simulator can be configured with an exact geolocation, the actual receiver output was compared against the true position, and the Euclidean RMS distance error was calculated. For one scenario, and since navigation doesn't require online operation, the same offline recording file used in acquisition and tracking results was used. It contains 12 satellites and a duration of 5 minutes. Both receivers were set with similar parameters for acquisition, tracking, and navigation, such as: sampling rate of 5 MHz, 4 msec coherent integration length, 10 KHz acquisition search band, 10 samples navigation averaging window, 5 Hz navigation sampling output rate, among others. For a second and third scenarios, real GPS signal recordings were used, which were recorded in an exact same location in Colorado Springs, CO during a static test. These two scenarios were recorded with OCXO, and with OCXO + GPSDO. These tests address receiver performance with real signals, and at the same time with and without GPS disciplining (GPSDO). These tests no. 2 and no. 3 do not have true mean error, since a true geolocation was not obtained.

Table XV shows both RMS error and true mean error results for proposed receiver and *GNSS-SDR*. For the true mean error, *GNSS-SDR* showed a bias error in height of 12m on average.

TABLE XV
RECEIVER NAVIGATION PVT PRECISION COMPARISON ON HOST PC

| SDR PVT Precision Scenarios | Scenario | RMS Mean Error (m) | True Mean Error (m) |
|---|---|---|---|
| Proposed SDR | 1 (SIM) | 0.62 | 1.33 |
|  | 2 (OCXO) | 10.23 | N/A |
|  | 3 (GPSDO) | 4.28 | N/A |
| *GNSS-SDR* | 1 (SIM) | 1.14 | 11.99 |
|  | 2 (OCXO) | N/A | N/A |
|  | 3 (GPSDO) | 19.34 | N/A |

* Scenario 1 comes from GPS simulator in a static environment and known geolocation.
* Scenarios 2 and 3 are from a real GPS static location in Colorado Springs, CO, using the OCXO with and without disciplining, i.e. OCXO, and OCXO + GPSDO.

For RMS error, both receivers performed well, with a slight advantage on the proposed SDR. As for tests 2 and 3, proposed SDR showed a clear precision improvement of more than 50% when using GPS disciplining from the board mounted kit, as opposed to only OCXO performance, as discussed in section III-A. As for *GNSS-SDR*, the receiver showed a RMS error of 19.34m for the disciplined scenario, and for the OCXO only, the receiver was not able to acquire the signal properly. The comparison of both receiver performances with real GPS signals and GPSDO enabled (scenario 3) showed a gain of more than 4 times in precision on the reference receiver. Preliminary results show visual stability of the navigation module of the receiver in Fig. 8 where a 8-min drive test in Colorado Springs, CO, was conducted.

## VI. CONCLUSION

This paper proposed and demonstrated a new generation state-of-the-art GPS receiver based on LV and C/C++ integrations as DLL modules. Several acceleration factors for SDR were discussed and leveraged in proposed SDR for real-time operation. Conventional baseband modules for acquisition, tracking, and navigation were used as DLL modules integrated into LV, as well as a USRP B200 front-end paired with an Intel NUC5i5RYK as the receiver hardware testbed were evaluated.

An extensive comparative analysis was made with existing open-source SDR solutions in both offline and online mode of operation. For all three baseband modules: acquisition, tracking, and navigation, several comparisons were made against acceleration factors, and other receivers such as *fast-gps* and *GNSS-SDR*. Four versions of the proposed receiver were considered based on the acceleration factors discussed: Google Maps, multi-producer/consumer loop, Parallelizable loops, and SIMD features. The paper also presented an approach to develop an advanced receiver to address various receiver tasks including multipath measurements as reported in [9]. Different from [9] the real-time receiver functionality is achieved by using built-in features of the selected environment. The paper studied the impact of various acceleration features and their practical implementation intricacies.

For acquisition, an advanced joint-search FFT acquisition algorithm [26] was tested in proposed receiver against the conventional PCS acquisition algorithm. Optimization libraries for FFT operations were also compared. Both LV-based receiver and CLI-only receiver were compared to assess any overhead by former receiver. For tracking, several online and offline tests were evaluated. Tests involving quantization carrier wave performance and relative loss, receiver versions performance gains, and against *GNSS-SDR* were compared. Metrics such as CPU load and memory occupancy were assessed for online operation of receivers. A maximum of 8 real-time channel tracking at 25 MHz sampling rate was achieved on latest version of proposed receiver. For navigation, PVT solution precision was assessed for proposed SDR and *GNSS-SDR*.

Eventually, we present strengths as well as weaknesses of the proposed solution, such as relying on certain platform and thus limiting high-end optimizations for advanced commercial-grade solutions. The proposed solution promises fast prototyping for determined research purposes and achieves computational space for future plans.

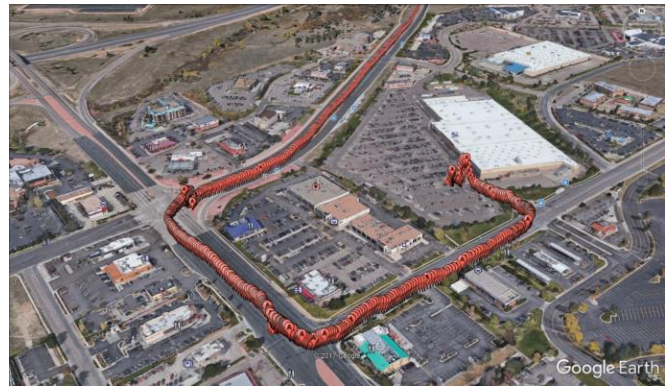

Fig. 8. Preliminary results for dynamic test in Colorado Springs, CO with OCXO + GPSDO mode.

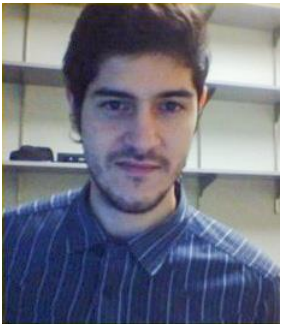

**Erick Schmidt** (S'17) received the B.Sc. degree in electronics and computer engineering with Honors from The Monterrey Institute of Technology and Higher Education, Monterrey, Mexico, in 2011, and the M.Sc. degree from the University of Texas at San Antonio, San Antonio, TX, USA, in 2015. From 2011 to 2013, he was a Systems Engineer with Qualcomm Incorporated.

He is currently working towards the Ph.D. degree in electrical engineering from the University of Texas at San Antonio, San Antonio, TX, USA. His research interests include software-defined radio, indoor navigation, global navigation satellite system (GNSS), and fast prototyping algorithms and accelerators for baseband communication systems. He is a student member of the IEEE and the Institute of Navigation (ION).

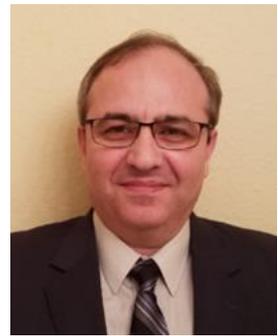

**David Akopian** (M'02-SM'04) is a Professor at the University of Texas at San Antonio (UTSA). Prior to joining UTSA, he was a Senior Research Engineer and Specialist with Nokia Corporation from 1999 to 2003. From 1993 to 1999 he was a researcher and instructor at the Tampere University of Technology, Finland, where he received his Ph.D. degree in 1997. Dr. Akopian's current research interests include digital signal processing algorithms for communication and navigation receivers, positioning, dedicated hardware architectures and platforms for software defined radio and communication technologies for healthcare applications.

He authored and co-authored more than 30 patents and 140 publications. He is elected as a Fellow of US National Academy of Inventors in 2016. He served in organizing and program committees of many IEEE conferences and co-chairs an annual conference on Multimedia and Mobile Devices. His research has been supported by National Science Foundation, National Institutes of Health, USAF, US Navy, and Texas foundations.

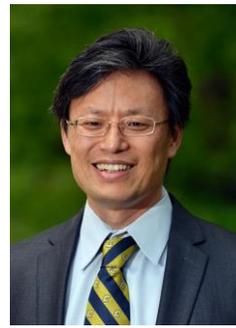

**Daniel J. Pack** received the Bachelor of Science degree in electrical engineering from Arizona State University, Tempe, AZ, USA, in 1988, the Master of Science degree in engineering sciences from Harvard University, Cambridge, MA, USA, in 1990, and the Ph.D. degree in electrical engineering from Purdue University, West Lafayette, IN, USA, in 1995.

He is currently the Dean of the College of Engineering and Computer Science, University of Tennessee, Chattanooga (UTC), TN, USA. Prior to joining UTC, he was a Professor and the Mary Lou Clarke Endowed Chair of the Electrical and Computer Engineering Department, University of Texas, San Antonio and a Professor (now Professor Emeritus) of electrical and computer engineering at the United States Air Force Academy, CO, USA, where he was the founding Director of the Academy Center for Unmanned Aircraft Systems Research. His research interests include unmanned aerial vehicles, intelligent control, automatic target recognition, robotics, and engineering education.